\documentclass[twocolumn,epjc3-upd]{svjour3}          

\RequirePackage[T1]{fontenc}
\RequirePackage{fix-cm}

\smartqed  

\RequirePackage{graphicx}
\RequirePackage{mathptmx}      
\RequirePackage{flushend}
\RequirePackage[numbers,sort&compress]{natbib}
\RequirePackage[colorlinks,citecolor=blue,urlcolor=blue,linkcolor=blue]{hyperref}

\usepackage[amssymb]{SIunits}
\usepackage{lineno}
\usepackage[pdftex,dvipsnames,hyperref]{xcolor}
\journalname{Eur. Phys. J. C}
\begin{document}
\title{Deep Sea Tests of a Prototype of the KM3NeT Digital Optical Module}

\subtitle{KM3NeT Collaboration}

\author{
S.~Adri\'{a}n-Mart\'{i}nez\thanksref{UPV} \and
M.~Ageron\thanksref{CPPM} \and
F.~Aharonian\thanksref{DIAS} \and
S.~Aiello\thanksref{I-CAT} \and
A.~Albert\thanksref{GRPHE} \and
F.~Ameli\thanksref{I-ROM} \and
E.G.~Anassontzis\thanksref{U-ATH} \and
M.~Anghinolfi\thanksref{I-GEN} \and
G.~Anton\thanksref{ECAP} \and
S.~Anvar\thanksref{C-SED}
% ...
}

\institute{
\label{UPV}{Instituto de Investigaci\'{o}n para la Gesti\'{o}n Integrada de las Zonas Costeras,Universitat Polit\`{e}cnica de Val\`{e}ncia,~Gandia,~Spain}
\and
\label{CPPM}CPPM,~Aix-Marseille Universit\'{e},~CNRS/IN2P3,~Marseille,~France
\and
\label{DIAS}DIAS, Dublin, Ireland
\and
\label{I-CAT}INFN, Sezione di Catania, Catania, Italy
\and
\label{GRPHE}GRPHE, Universit\'{e} de Haute Alsace, IUT de Colmar, Colmar, France
\and
\label{I-ROM}INFN, Sezione di Roma, Roma, Italy
\and
\label{U-ATH}National and Kapodistrian University of Athens, Deparment of Physics, Athens, Greece
\and
\label{I-GEN}INFN, Sezione di Genova, Genova, Italy
\and
\label{ECAP}Erlangen Centre for Astroparticle Physics, Friedrich-Alexander-Universit{\"a}t Erlangen-N{\"u}rnberg,Erlangen, Germany
\and
\label{C-SED}CEA, Irfu/Sedi, Centre de Saclay, Gif-sur-Yvette, France
\and
\label{I-NAP}INFN, Sezione di Napoli, Napoli, Italy
% ...
\and
\label{U-NAP}Universit\`{a} 'Federico II', Dipartimento di Fisica, Napoli, Italy
\and
\label{I-BAR}INFN, Sezione di Bari, Bari, Italy
% ...
\and
\label{I-PIS}INFN, Sezione di Pisa, Pisa, Italy 
\and
\label{U-PIS}Universit{\`a} di Pisa, Dipertimento di Fisica , Pisa, Italy
\and
\label{I-BOL}INFN, Sezione di Bologna, Bologna, Italy
\and
\label{U-BOL}Universit\'a di Bologna, Dipartimento di Fisica e Astronomia, Bologna, Italy
% ...
\and
\label{U-LEI}Leiden Institute of Physics, Leiden University, Leiden, The Netherlands
\and
\label{U-ROM}Universit{\`a} di Roma La Sapienza, Dipartimento di Fisica, Roma, Italy
\and
\label{U-SAL}Universit{\`a} di Salerno, Dipartimento di Fisica, Fisciano, Italy
\and
\label{U-UVA}Institute of Physics, University of Amsterdam, Amsterdam, The Netherlands
\and
\label{I-LNS}INFN, Laboratori Nazionali del Sud, Catania, Italy
\and
\label{U-PSUD}Universit\'{e} Paris-Sud , 91405 Orsay Cedex, France
% ...
\and
\label{U-THS}Aristotle University Thessaloniki, Thessaloniki, Greece
\and
\label{U-WZB}University W{\"u}rzburg, W{\"u}rzburg, Germany
\and
\label{I-LNF}INFN, INFN, Laboratori Nazionali di Frascati, Frascati, Italy
% ...
\and
\label{U-TUB}Eberhard Karls Universit{\"a}t T{\"u}bingen, T{\"u}bingen, Germany
\and
\label{U-UU}Utrecht University, Utrecht, The Netherlands
\and
\label{U-BAM}Dr. Remeis Sternwarte, Friedrich-Alexander-Universit{\"a}t Erlangen-N{\"u}rnberg, Bamberg, Germany
\and
\label{U-CAT}Universit\`{a} di Catania, Dipartimento di Fisica ed Astronomia, Catania, Italy
\and
\label{U-CYP}University of Cyprus, Physics Department, Nicosia, Cyprus
\and
\label{U-AEG}University of Aegean, Athens, Greece
% ...
\and
\label{U-ABD}Oceanlab, University of Aberdeen, Aberdeen, United Kingdom
\and
\label{U-GEN}Universit\`{a} di Genova, Dipartimento di Fisica, Genova, Italy
\and
\label{C-SPP}CEA, Irfu/SPP, Centre de Saclay, Gif-sur-Yvette, France
\and
\label{U-SHF}University of Sheffield, Department of Physics and Astronomy, Sheffield, United Kingdom
% ...
}

\thankstext{ACCNL}{Also at Accademia Novale di Livorno, Livorno, Italy}
\thankstext[$\star$]{ca}{Corresponding authors, creusot@in2p3.fr; h84@nikhef.nl}




\authorrunning{KM3Net Collaboration} 
\titlerunning{KM3NeT DOM}

\date{Received: date / Accepted: date}

\maketitle

\begin{abstract}
The first prototype of a photo-detection unit of the future KM3NeT
neutrino telescope has been deployed in the deep waters of the Mediterranean
Sea. This digital optical module has a novel design with a very large photocathode area segmented by the use of 31 three inch photomultiplier tubes. It has been integrated in the
 ANTARES detector for {\em in-situ} testing and validation. This paper 
reports on the first months of data taking and rate measurements. The analysis results highlight the capabilities of
the new module design in terms of background suppression and signal recognition. The directionality of the optical module enables the recognition of multiple Cherenkov photons from the same $^{40}$K decay and the localisation bioluminescent activity in the neighbourhood.
The single unit can cleanly identify atmospheric muons and provide sensitivity
to the muon arrival directions.
\end{abstract}
\begin{figure}[t]
  \centering
\setlength{\unitlength}{1pt}
\begin{picture}(250,380)
 \put(40,190) {\includegraphics[width=0.3\textwidth]{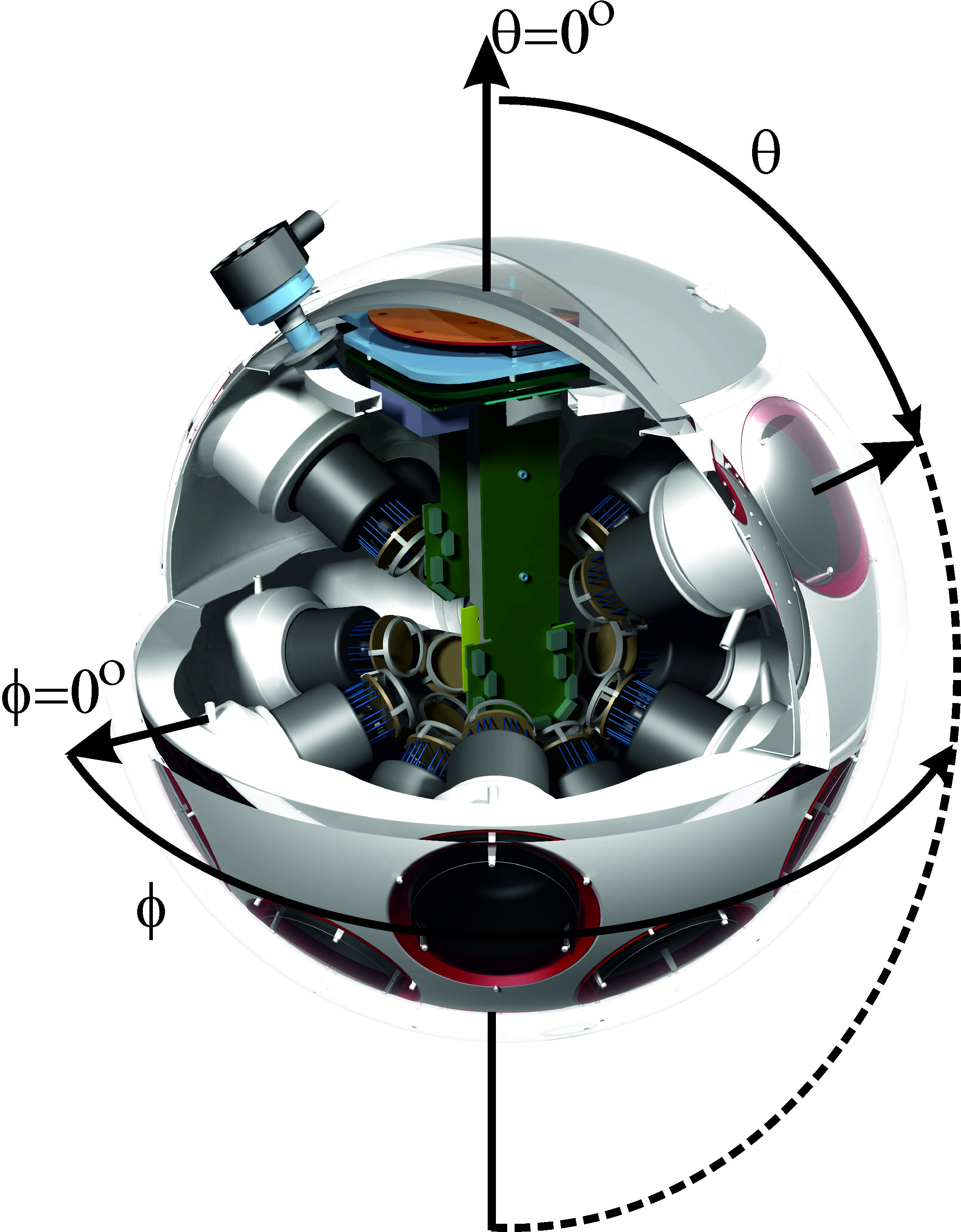}} 
\put(10,0) {  \includegraphics[width=0.45\textwidth]{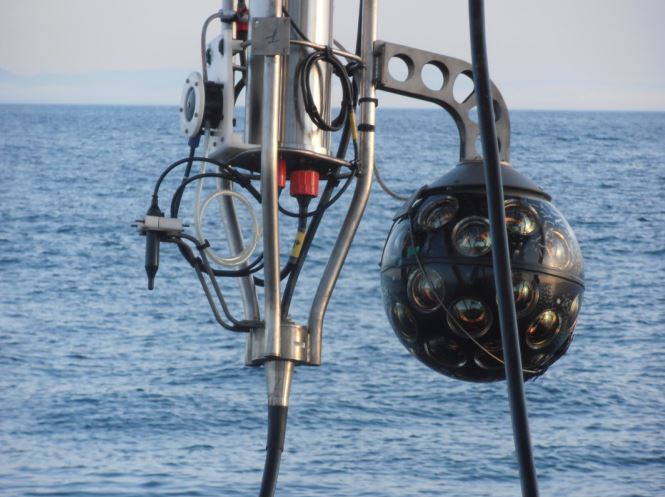}}
\put(200,350) { \makebox(0,0)[l]{\bf (A)} }
\put(200,150) { \makebox(0,0)[l]{\bf (B)} }
\end{picture}
  \caption{(A) A 3D cutout drawing of the DOM. The zenith angle $\theta$ and the azimuthal angle $\phi$ are indicated. Vertically upward corresponds to $\theta = 0\degree$ whereas $\phi = 0\degree$ points to the support cable.  (B) The DOM connected to the ANTARES instrumentation line. The structure to which the DOM is connected is a standard ANTARES support frame including a cylindrical electronics container.}\label{f:array}
\end{figure}
\section{Introduction}
\label{intro}
The KM3NeT Observatory~\cite{b:km3net} is a large scale neutrino telescope to be
built in the deep waters of the Mediterranean Sea. With several cubic
kilometres instrumented with thousands of optical
sensors, KM3NeT will be the largest and most sensitive high energy neutrino detector.
It will be capable of neutrino astronomy with unprecedented accuracy.
Being situated in the Northern Hemisphere it will be  particularly suited to the investigation of high energy neutrinos from our Galaxy. 
The IceCube collaboration has recently reported a first signal of 
 neutrinos with energies of PeVs~\cite{b:iceCube1}. A subsequent analysis showed indeed an excess of several tens
of events that are attributed to extraterrestrial sources~\cite{b:iceCube2}. 
This signal with energies up to ~2 PeV strengthens the motivation
 for the construction of KM3NeT. 
Several types of astrophysical objects have been proposed as sites where hadrons are accelerated to extreme energies.
The interaction of these particles with matter or radiation  in or near the source produces pions and subsequently 
high energy neutrinos~\cite{b:nuProd1,b:nuProd2,b:nuProd3}. These neutrinos propagate with almost no interaction and may reach the Earth undisturbed. From the observed neutrino direction the 
sources may be identified~\cite{b:diffuse,b:pointsource}.
\\The neutrino detection is based on sampling the Cherenkov
light induced by the particles produced in a  neutrino
interaction in the  vicinity of the detector. The muon produced in the 
charged current interaction of a muon-neutrino provides through its long range in water
a particularly sensitive detection channel, but the huge volume of KM3NeT also provides large sensitivity to the other neutrino flavours and to 
the neutral current interactions. The properties of the deep Mediterranean water allow for high accuracy in the determination of the neutrino direction. 
This has been demonstrated in the KM3NeT predecessor ANTARES~\cite{b:ANTpointsource}.\\
The sampling of Cherenkov light in the KM3NeT telescope is performed with
the Digital Optical Modules (DOMs). A prototype of these is the subject of this paper. It has been installed on the instrumentation line of the ANTARES detector~\cite{b:ant1}. 
A technical drawing of the DOM and a picture of the DOM connected to the ANTARES line during deployment are shown in 
figure~\ref{f:array}. 
\begin{figure}[t]
  \centering
\setlength{\unitlength}{1pt}
\begin{picture}(250,300)
\put(0,150) { \includegraphics[width=0.40\textwidth]{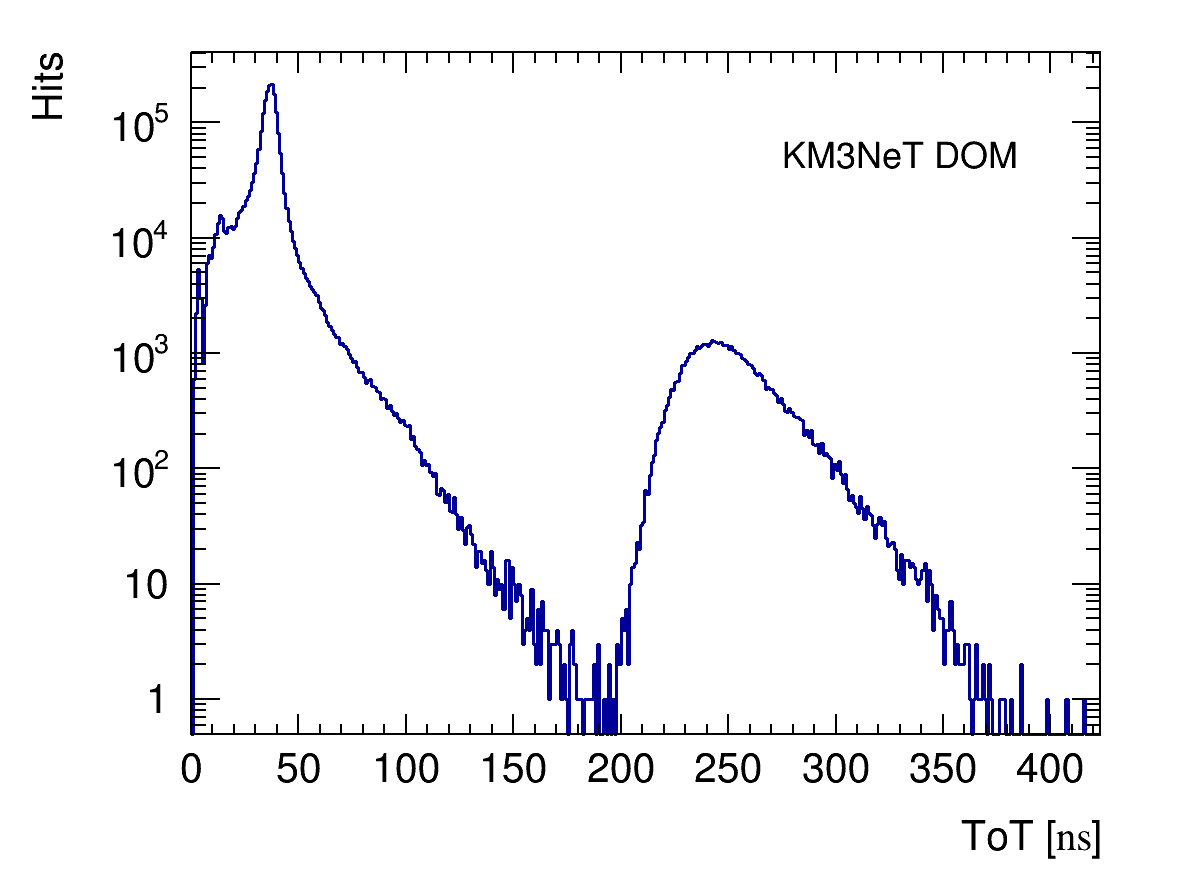}}
\put(0,0){ \includegraphics[width=0.40\textwidth]{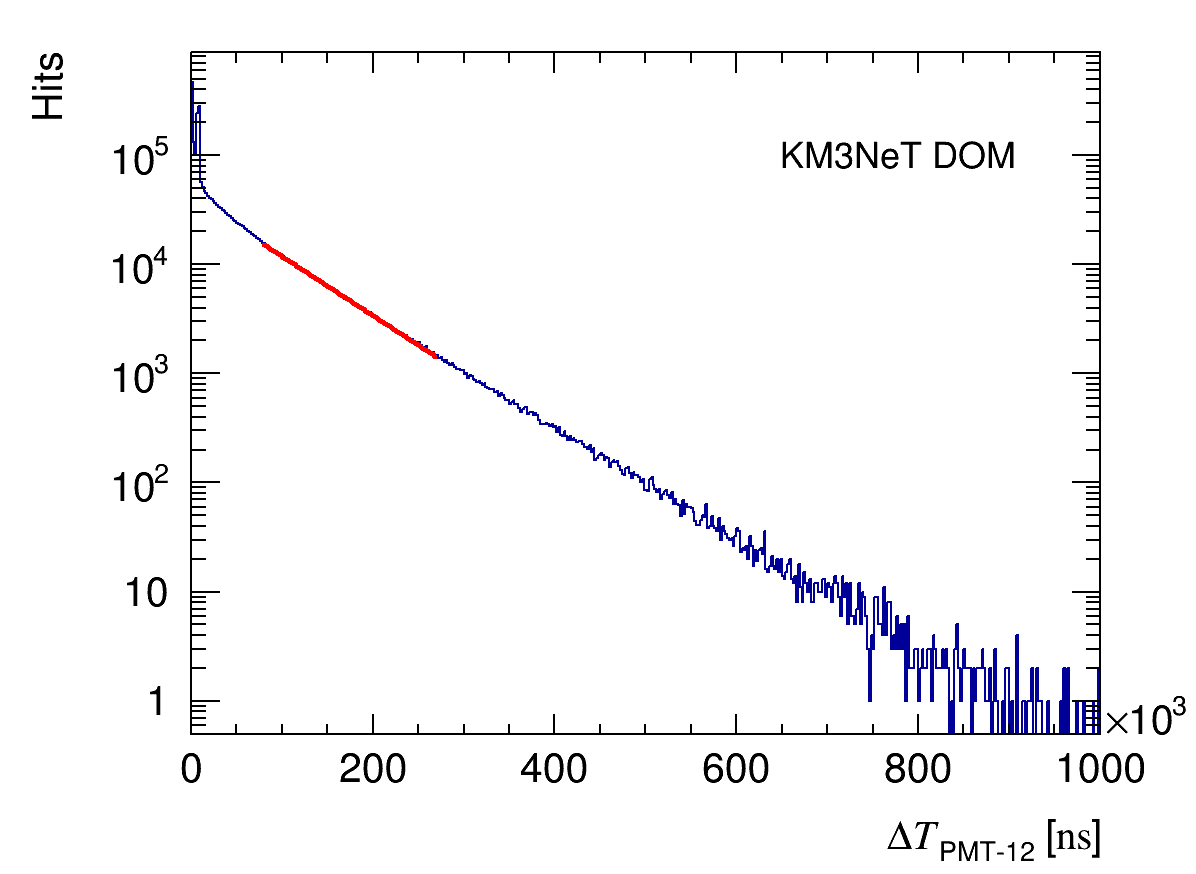}} 
\put(140,253) { \makebox(0,0)[l]{\bf (A)} }
\put(140,103) { \makebox(0,0)[l]{\bf \bf (B)} }
\end{picture}
   \caption{ (A) Numbers of hits as a function of ToT for one of the PMTs. The broad ToT peak corrresponds to the large pulse height
    caused by the laser calibration. (B) Histogram of the time
    difference between two consecutive hits. The exponential fit, as indicated by the red line,  gives a characteristic time $\tau =
    \unit{1.05\times 10^{-4}}{s}$.}\label{f:firstrun}
\end{figure}

\section{Digital Optical Module Design and Deployment}
\label{s:detDes}
The  DOM consists of a glass sphere\footnote{Vitrovex, NAUTILUS Marine Service GmbH, Buxtehude, Germany; http://www.nautilus-gmbh.com/vitrovex-deep-sea-housings}, $\unit{432}{mm}$ in
diameter.  The sphere houses 31  photomultiplier tubes (PMTs) with a photocathode diameter of
$\unit{72}{mm}$~\cite{b:km3net}. Each tube is surrounded by a cone-shaped reflector that effectively increases the diameter to about \unit{85}{mm}~\cite{b:oksana}.
 For the present prototype module, ETEL\footnote{ET Enterprises Ltd, Uxbridge, UB8 2YF, United Kingdom http://www.et-enterprises.com} D783FL PMTs are used.
 They are arranged in 5 rings of
 PMTs with  zenith angles  of 56\degree, 72\degree, 107\degree, 123\degree and 148\degree, respectively (see figure~\ref{f:array}A).
In each ring the 6 PMTs are spaced at 60\degree\  in azimuth and successive rings are staggered by 30\degree.
The last PMT points vertically downward at a zenith angle of 180\degree\ . \\ Each PMT has its own individual very-low-power
high-voltage base~\cite{b:timmer} with integrated amplification and adjustable discrimination. A multi-purpose multi-channel TDC incorporated within an FPGA was developed during the design phase of KM3NeT~\cite{b:zwart}. A modified version, implemented in the DOM, digitises the arrival time and the width of the discriminated PMT pulse, the time-over-threshold (ToT)~\cite{b:shebli}.
For the results presented in this paper, the threshold is set at the level of 0.3 of the mean single photon pulse height and the 
high voltage is set to provide an amplification of $3\times 10^6$. The data are transported the \unit{45}{km} to shore through the ANTARES \unit{1}{Gb/s}  multiplexed optical link, that uses reflective modulation in the DOM~\cite{b:jelle,b:tassos}.\\
As compared to traditional optical modules with single large PMTs~\cite{b:AntOM, b:NestOM,b:NemOM}, the design of the DOM has the advantage  that it houses three to four times the photocathode area 
in a single glass sphere and has an almost uniform angular coverage. Because the photocathode is segmented, the 
arrival of more than one photon  at the DOM is identified with  high efficiency and purity and provides a sensitivity to the direction of the detected light.
This allows for a highly efficient rejection of optical background. The number of pressure resistant vessels is minimised. Because of the low anode current the PMTs are expected to experience little ageing. In addition failure of a single PMT leaves the DOM still 97\% efficient. \\
The present DOM is the first KM3NeT optical module deployed and operated
in the deep sea. For this prototype step, the module was deployed as a
complete stand-alone detector on the instrumentation line of ANTARES. This line provides the DOM with power
and the optical connection to shore. On the 16$^{th}$ of April 2013 the line with the DOM attached was deployed and
connected using a ROV (Remotely Operated submersible Vehicle).
The line is situated 
on the  ANTARES detector site, south of Toulon, France (42\degree 50$^\prime$~N, 6\degree 10$^\prime$~E) at a depth of $\unit{2475}{m}$. Being connected to the line approximately \unit{100}{m} above the anchor, the DOM  operates at a depth of 
$\unit{2375}{m}$. Data taking commenced the day of deployment.
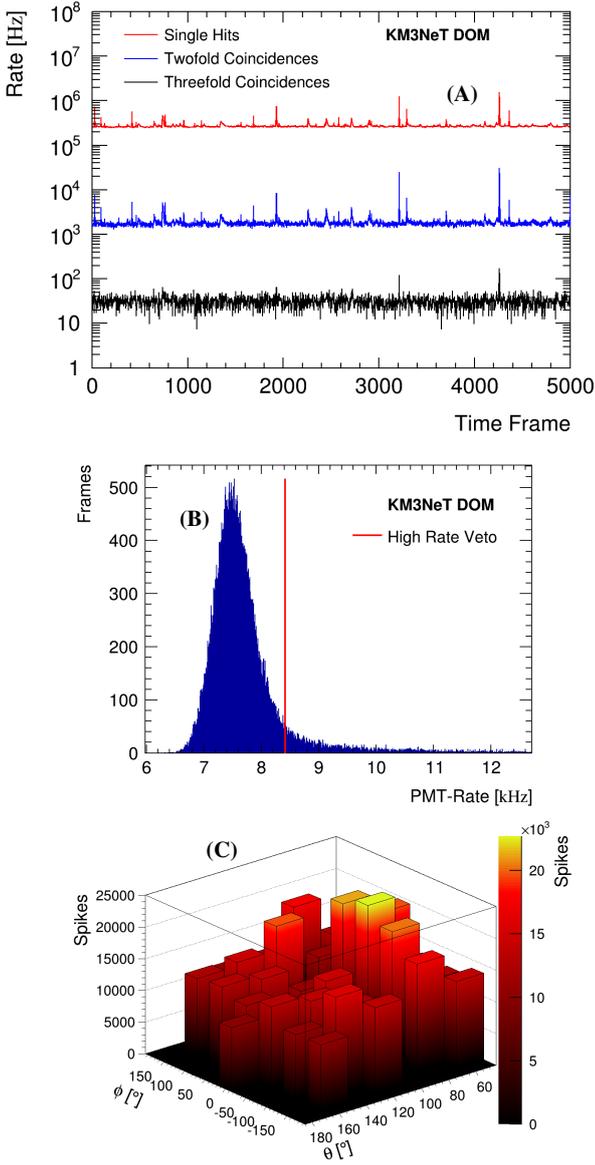
\begin{figure}[tp]
  \centering
\setlength{\unitlength}{1pt}
\begin{picture}(250,445)
 \put(0,280) { \includegraphics[width=235\unitlength]{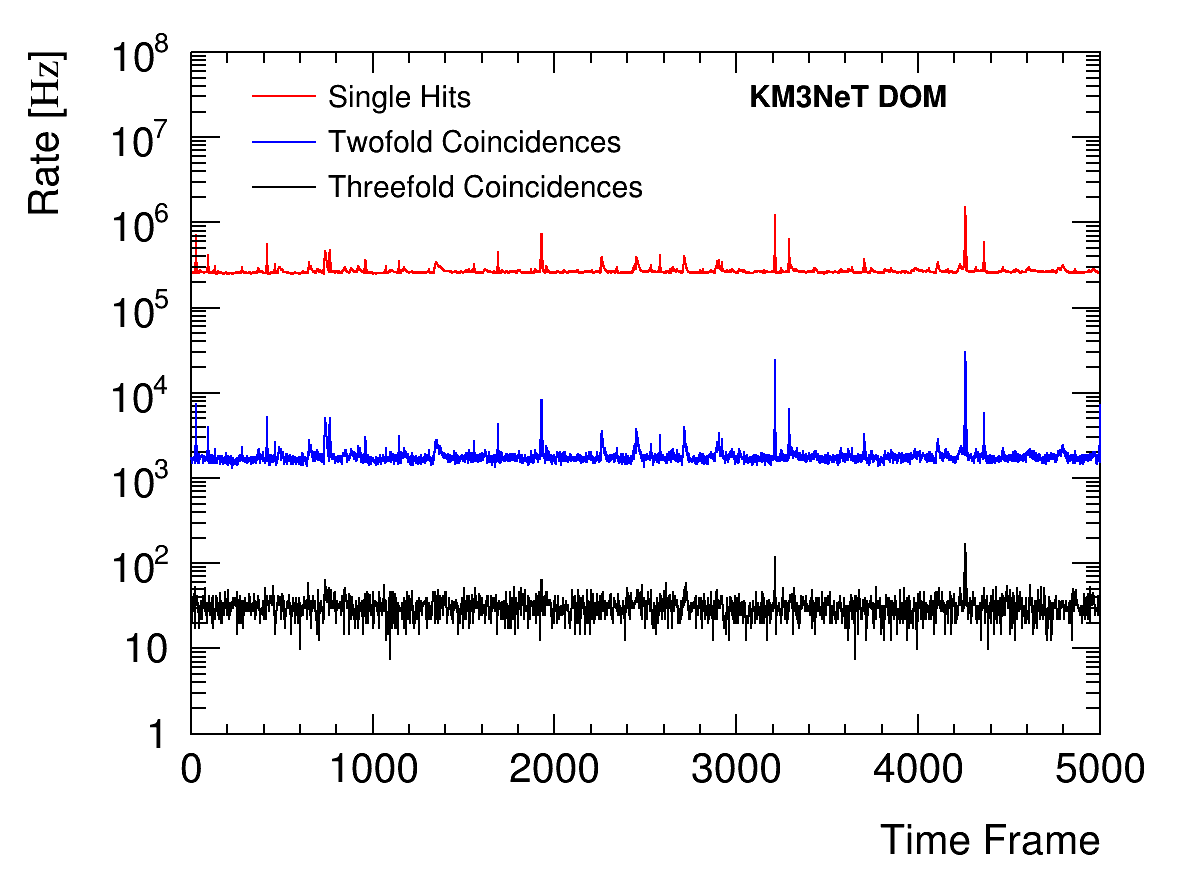}} 
\put(29,140) {\includegraphics[width=190\unitlength]{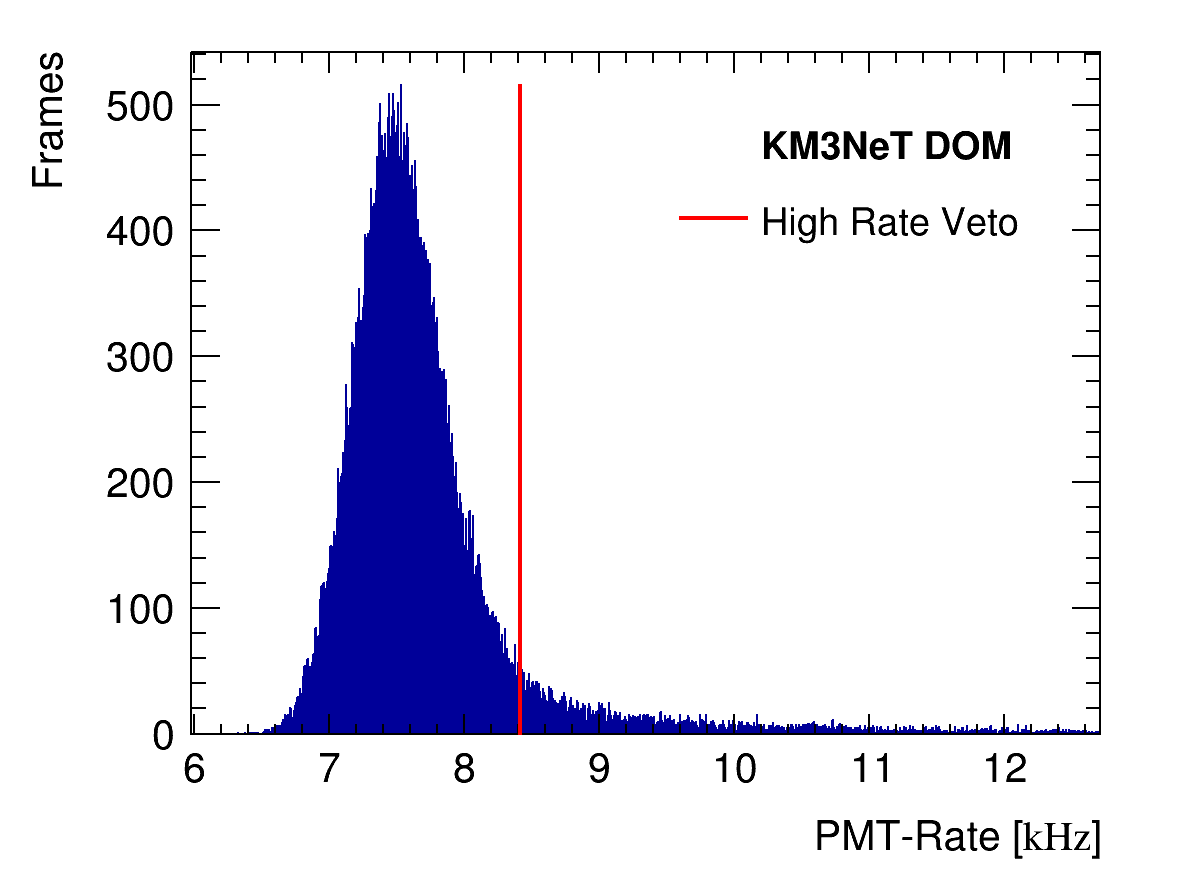}}
\put(29,0) {\includegraphics[width=190\unitlength]{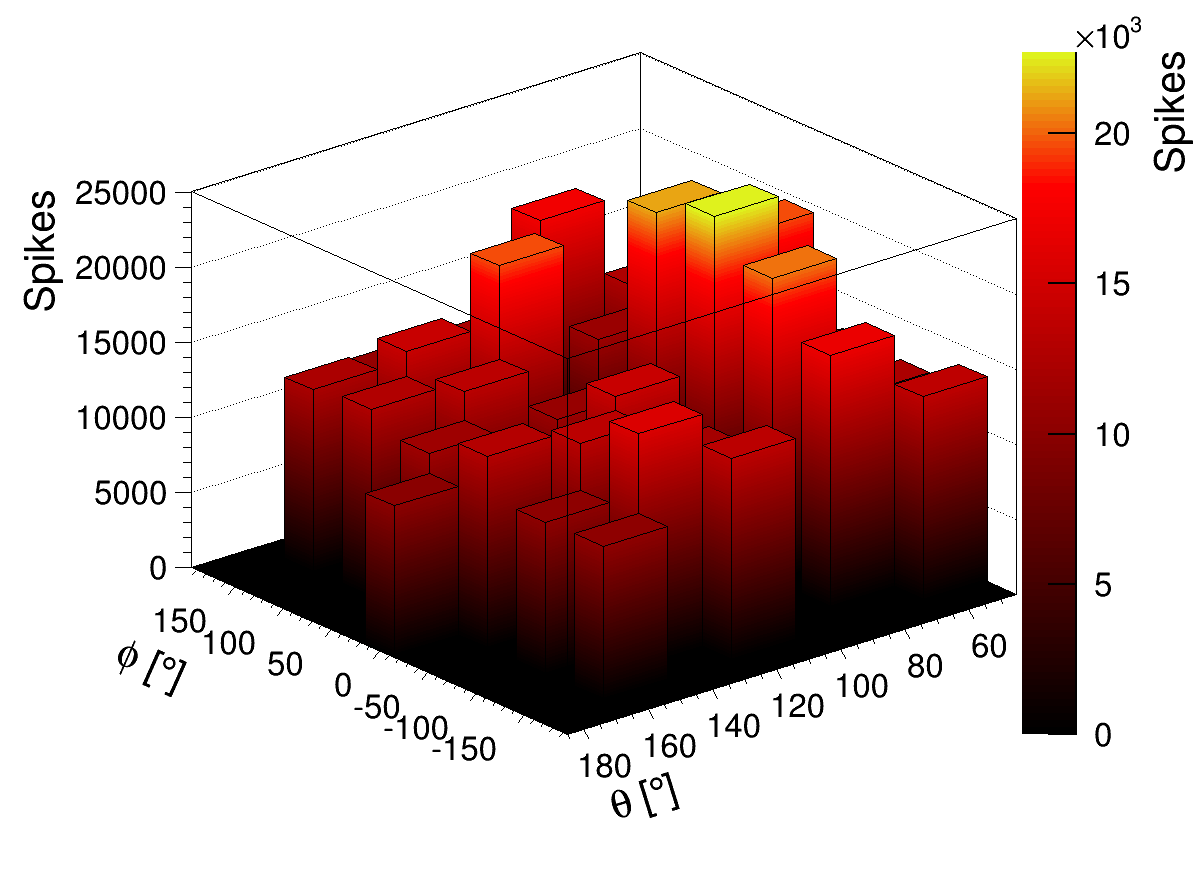}}
\put(170,410) { \makebox(0,0)[l]{\bf (A)} }
\put(70,250) { \makebox(0,0)[l]{\bf (B)} }
\put(80,125) { \makebox(0,0)[l]{\bf (C)} }
\end{picture}
\caption{(A) Aggregate hit rates as a function of the time measured in $\sim$\unit{134}{ms} bins. The top trace is for single hits, while the lower traces are for two- and threefold coincidences within a 20 ns window. (B) Histogram of the rate measured in \unit{134}{ms} timeframe bins for a representative PMT. The rising edge of the histogram is fit to a Gaussian  and the cutoff used in the further analysis is set at the mean plus 3 times the standard deviation (indicated by the red line).The cut-off is determined for each PMT separately. (C) The rate of bioluminescent bursts as a function of azimuth and zenith angles. The support structure is viewed by PMTs at small $\theta$ and near $\phi = 0\degree$.}\label{f:ratesJ}
\end{figure}
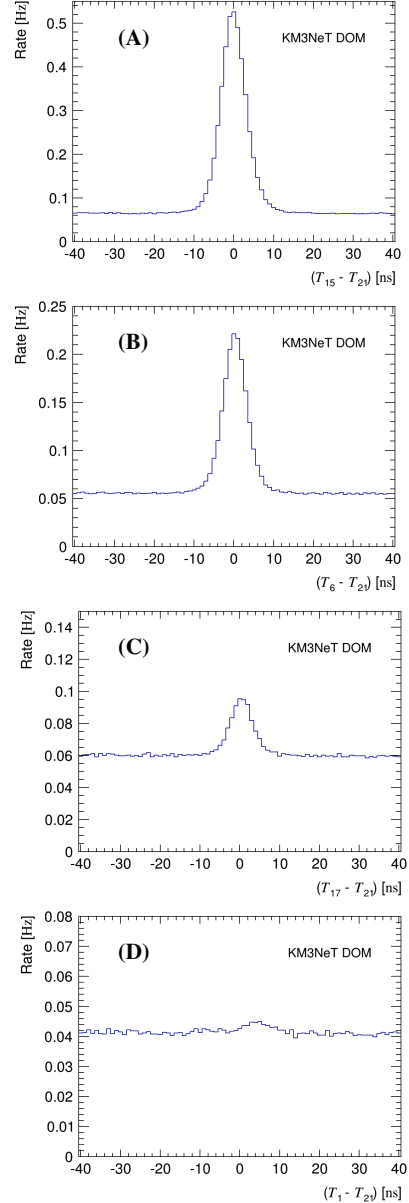
\begin{figure}[t]
  \centering
 \setlength{\unitlength}{1pt}
\begin{picture}(250,455)
 \put(40,345) {\includegraphics[width=0.32\textwidth]{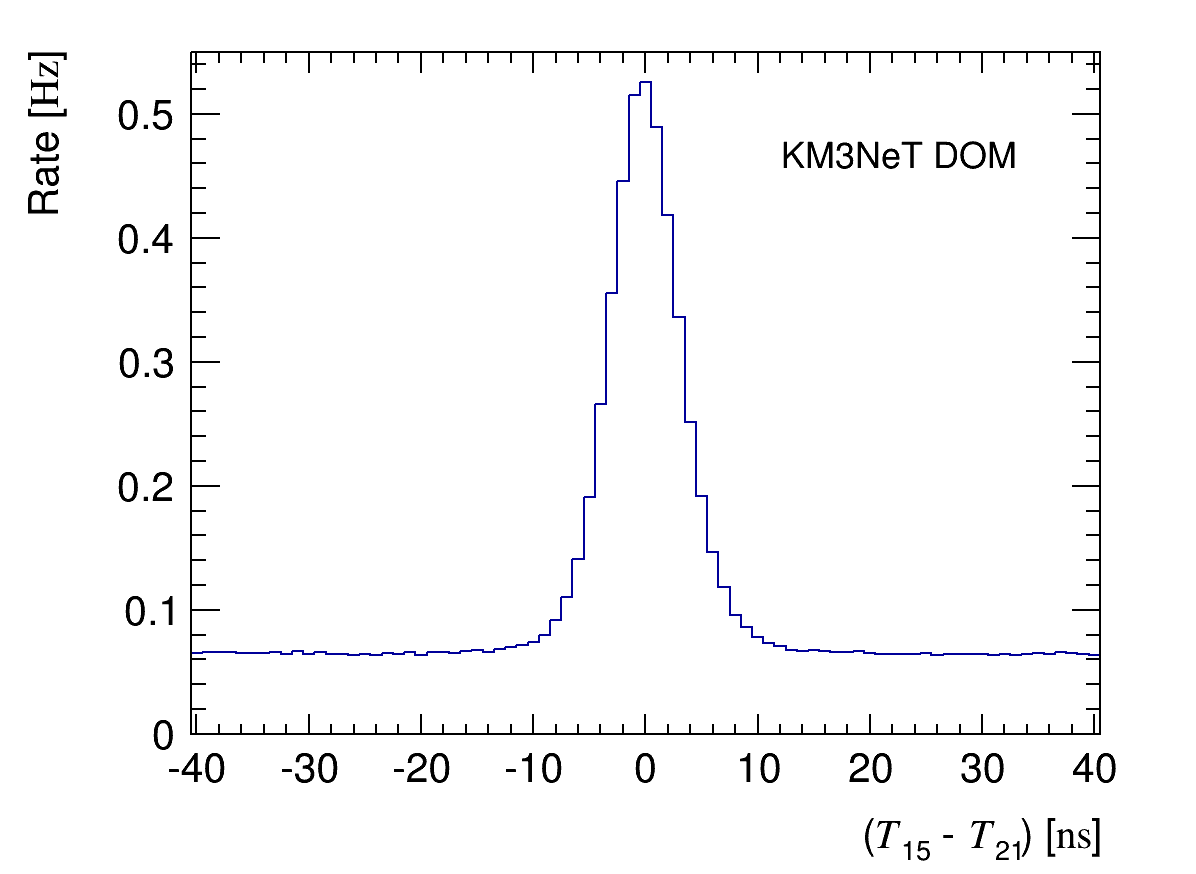}} 
\put(40,230) {\includegraphics[width=0.32\textwidth]{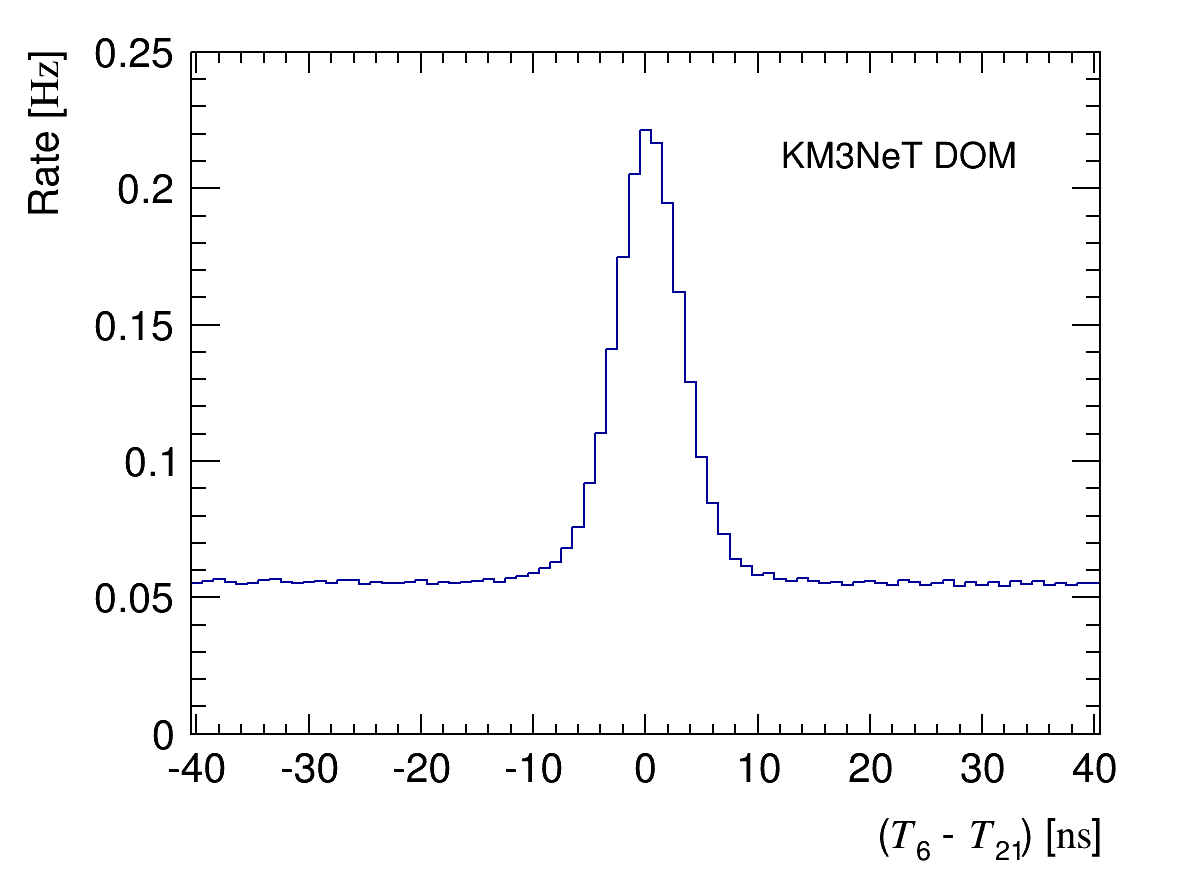}}
\put(40,115){ \includegraphics[width=0.32\textwidth]{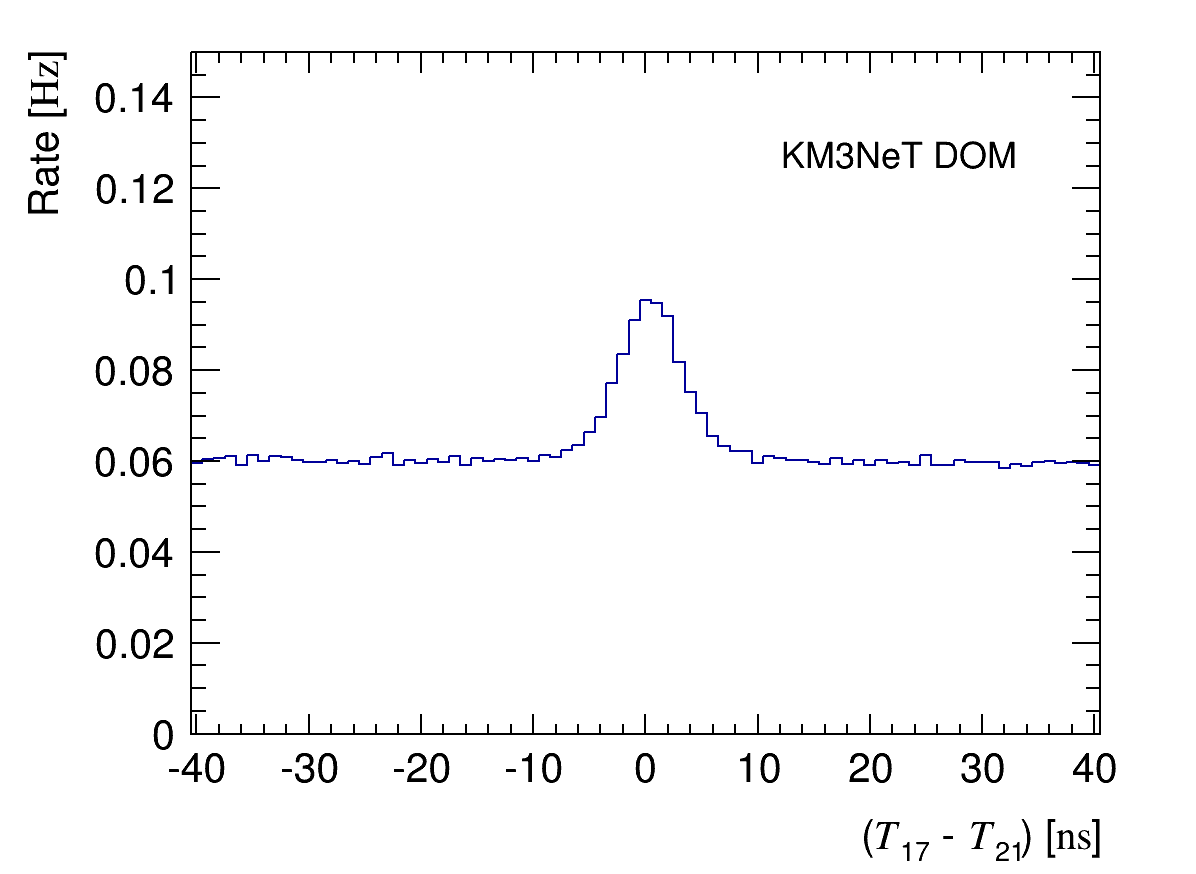}} 
\put(40,0){ \includegraphics[width=0.32\textwidth]{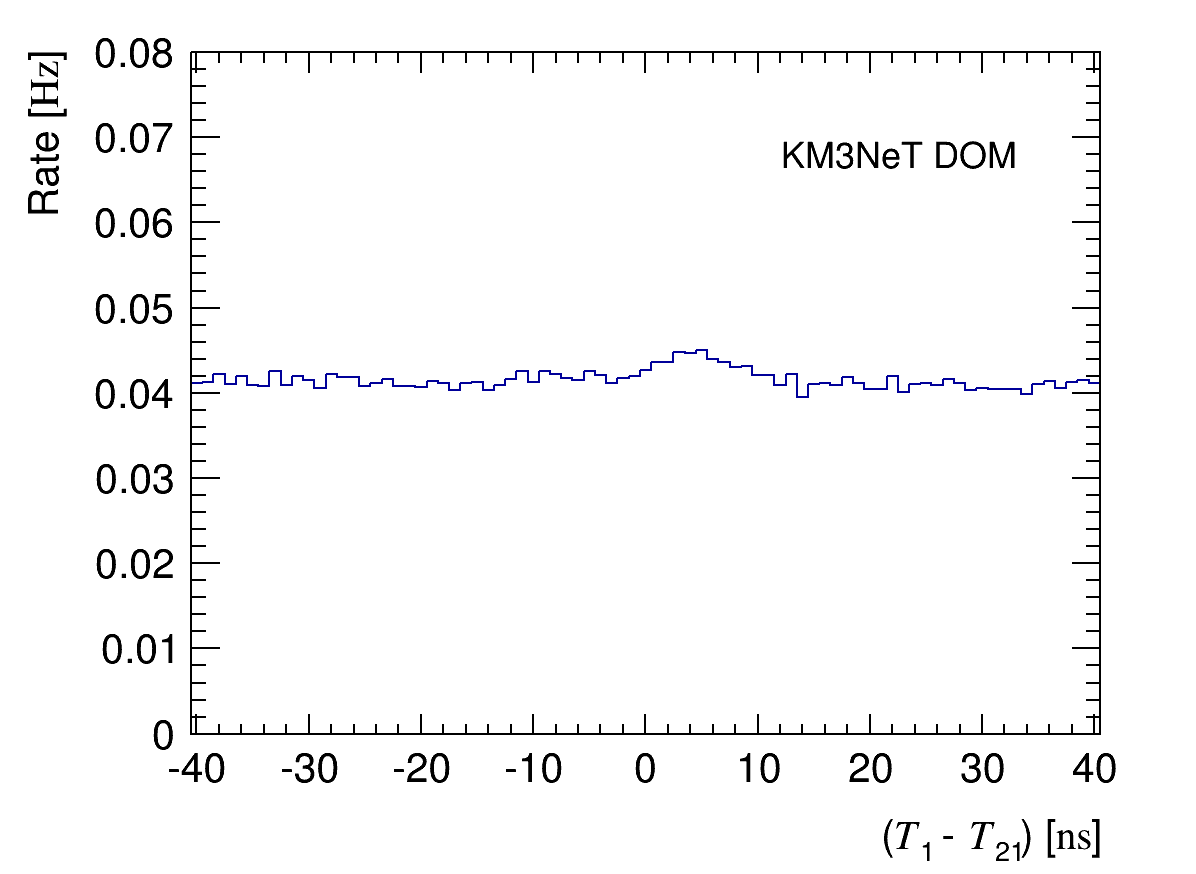}} 
\put(80,440) { \makebox(0,0)[l]{\bf (A)} }
\put(80,325) { \makebox(0,0)[l]{\bf (B)} }
\put(80,210) { \makebox(0,0)[l]{\bf (C)} }
\put(80,95) { \makebox(0,0)[l]{\bf (D)} }
\end{picture}
 \caption{Distribution of time differences between two PMTs of the DOM, with  an angular separation of (A) 33\degree , 
(B) 65\degree, (C) 120\degree\ and (D) 165\degree. }\label{f:times}
\end{figure}
\begin{figure}[t]
  \centering
 \setlength{\unitlength}{1pt}
\begin{picture}(250,175)
\put(-10,-5){ \includegraphics[width=0.50\textwidth]{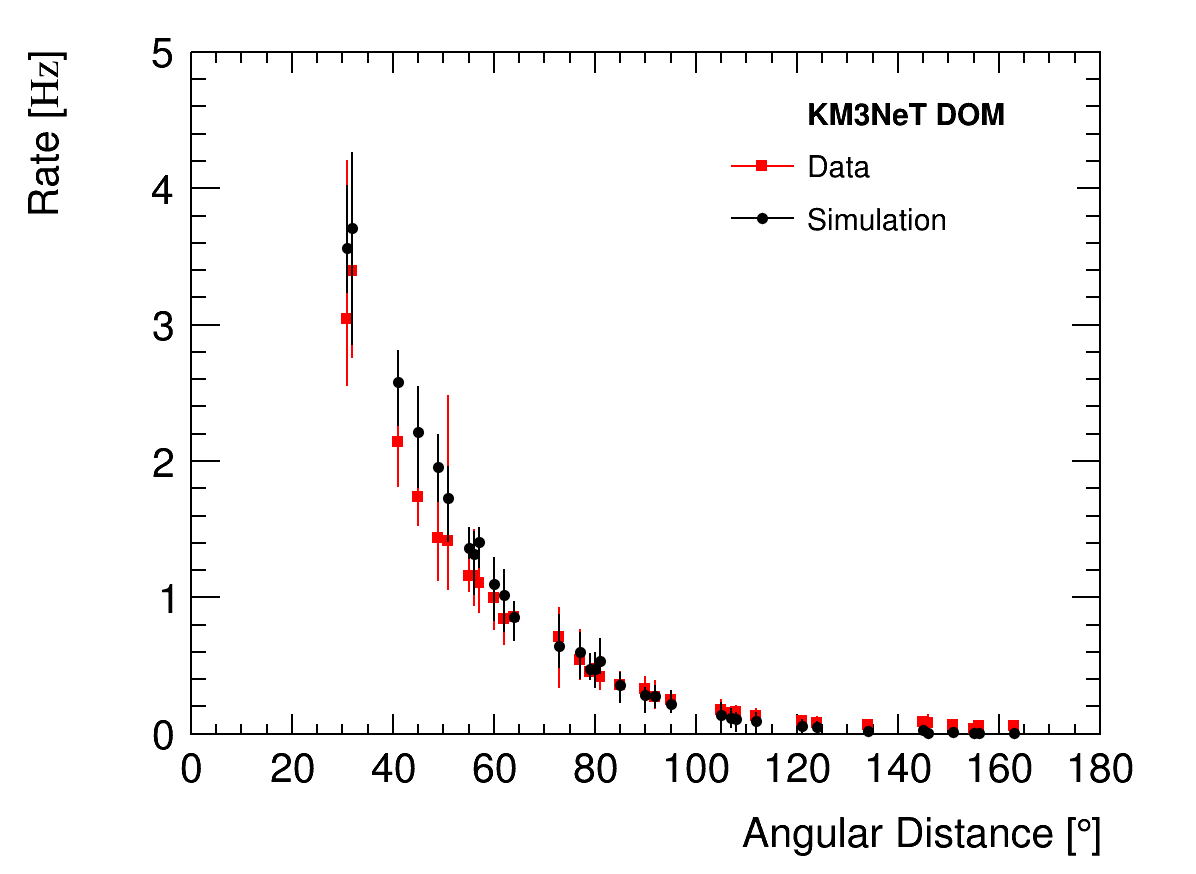} } 
\end{picture}
 \caption{The rate as a function of angular separation of the fitted coincidence signal is shown as a function of separation for both data (red squares) and simulation (black dots).}\label{f:times2}
\end{figure}

\section{Calibration}
\label{s:measur1}
The first data-taking runs were used to perform calibrations of the system and determine the 
counting rates of the individual PMTs. Figure~\ref{f:firstrun}A shows the result 
of one such calibration run. The graph shows the distribution of the registered ToT for one of the 
PMTs in the DOM. Two distinct peaks can be identified: one with a ToT attributable 
to single photon signals at \unit{35}{ns}, with a spread of  \unit{5.5}{ns} FWHM,  and one with significantly longer ToT caused by the flashing of the ANTARES calibration laser~\cite{b:ant2}.  Figure~\ref{f:firstrun}B gives the distribution of the time difference between successive hits, $\Delta T$, 
on a typical single PMT, showing the expected exponential shape. It has a slope $\tau = \unit{1.05\times 10^{-4}}{s}$
corresponding to a purely random background singles rate of around \unit{9.5}{kHz}, mostly due to $^{40}$K (see below). Because of its low repetition frequency of
\unit{1}{kHz} the signals from the laser do not appear in this plot. The increase in the rate near $\Delta T = 0$ is due to the approximately 5\% probability that these PMTs produce an afterpulse at around 3~$\mu$s after the initial pulse.\\
The aggregate hit rates of all PMTs in the DOM are shown in figure~\ref{f:ratesJ}A as a function of the time in \unit{134}{ms} timeframe bins,  for a run without laser. The baseline corresponds to an average rate of about $\unit{8}{kHz}$ per PMT and is  stable. Some timeframe bins show significant increases in the count rate. These increases can be attributed to bioluminescent activity.\\
Because the DOM contains many PMTs it is possible to look for coincidences of hits within the single optical module. Figure~\ref{f:ratesJ}A also shows the rate for twofold and threefold coincidences as a function of time. A two- (three-) fold coincidence is defined by the occurence of a hit in two (three) PMTs within a time window of 20 ns. To provide adequate statistics, the rate of threefold coincidences is averaged over three bins. As expected the twofold coincidences exhibit increased rates in the same bins due to the enhanced random coincidence rate. In the threefold coincidences, these increases are significantly diminished. The peaks in the  singles rates consist mainly of rate increases of a few PMTs at a time. The activity is not uniformly distributed over all PMTs but shows a prefered direction. A bioluminescent burst of a PMT is defined as a frame that exceeds the high-rate veto cut defined in Figure~\ref{f:ratesJ}B. If the preceding frame also fails the cut this is considered as the continuation of a single burst.  Figure~\ref{f:ratesJ}C shows the occurence rate of bioluminescent bursts as a function of the PMT position in zenith and azimuth, indicating that a significant amount of the bioluminescent activity is emanating from the direction of the support structure and electronics container. This type of structure will be absent in the final KM3NeT string, where the DOMs are supported by two thin, \unit{4}{mm} diameter ropes.  Averaged over the period July to December the rate of bursts per PMT is around one every 20 seconds for most PMTs and for the ones facing the structure it is 2.5 times larger. 
To select a very clean sample of hits, in the following a somewhat rigorous approach has been adopted (high-rate veto). If in any bin a single PMT shows a high rate activity, as defined in Figure~\ref{f:ratesJ}B, the bin is removed from the analysis.
Depending on the conditions this cut 
removes on average  10-15~\%  of the frames.\\ 
\begin{figure}[pt]
  \centering
 \setlength{\unitlength}{1pt}
\begin{picture}(250,600)
 \put(30,450) {\includegraphics[width=0.4\textwidth]{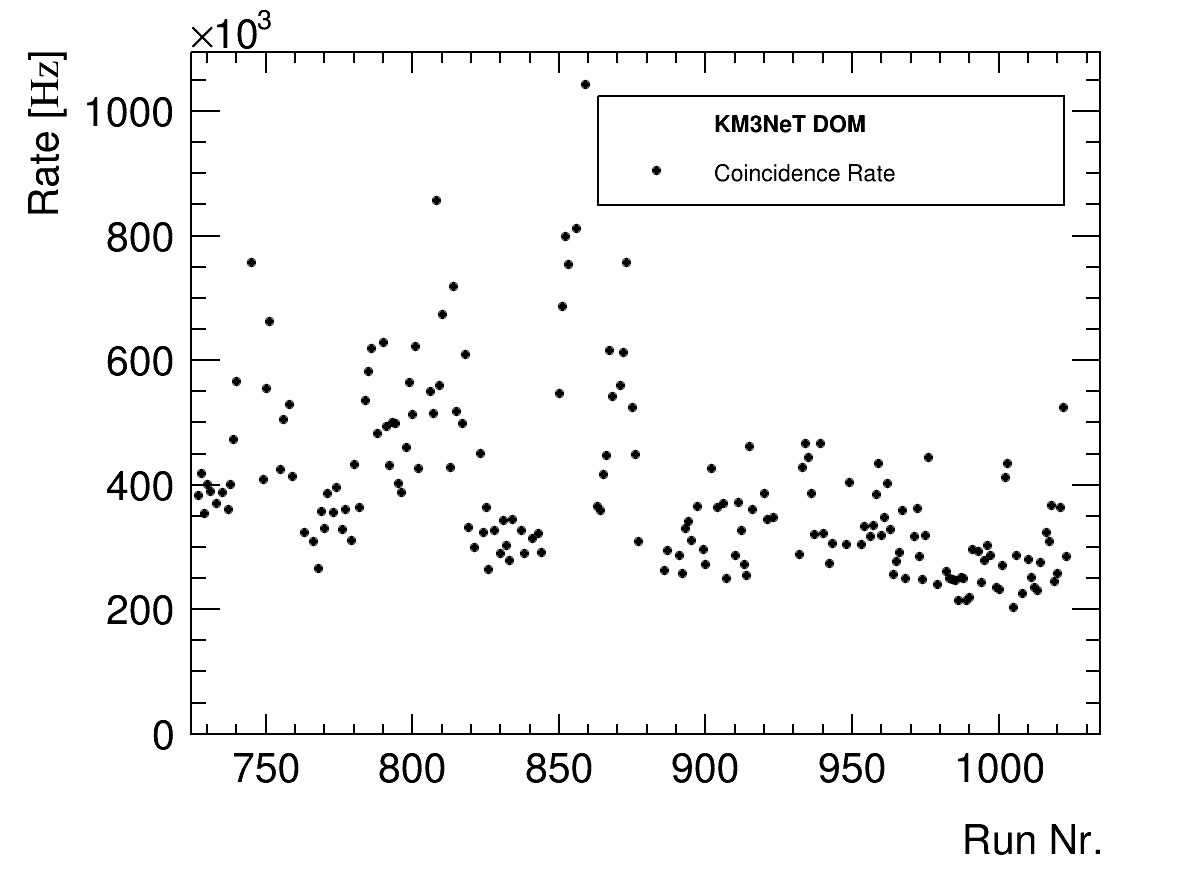}} 
\put(30,300) {\includegraphics[width=0.4\textwidth]{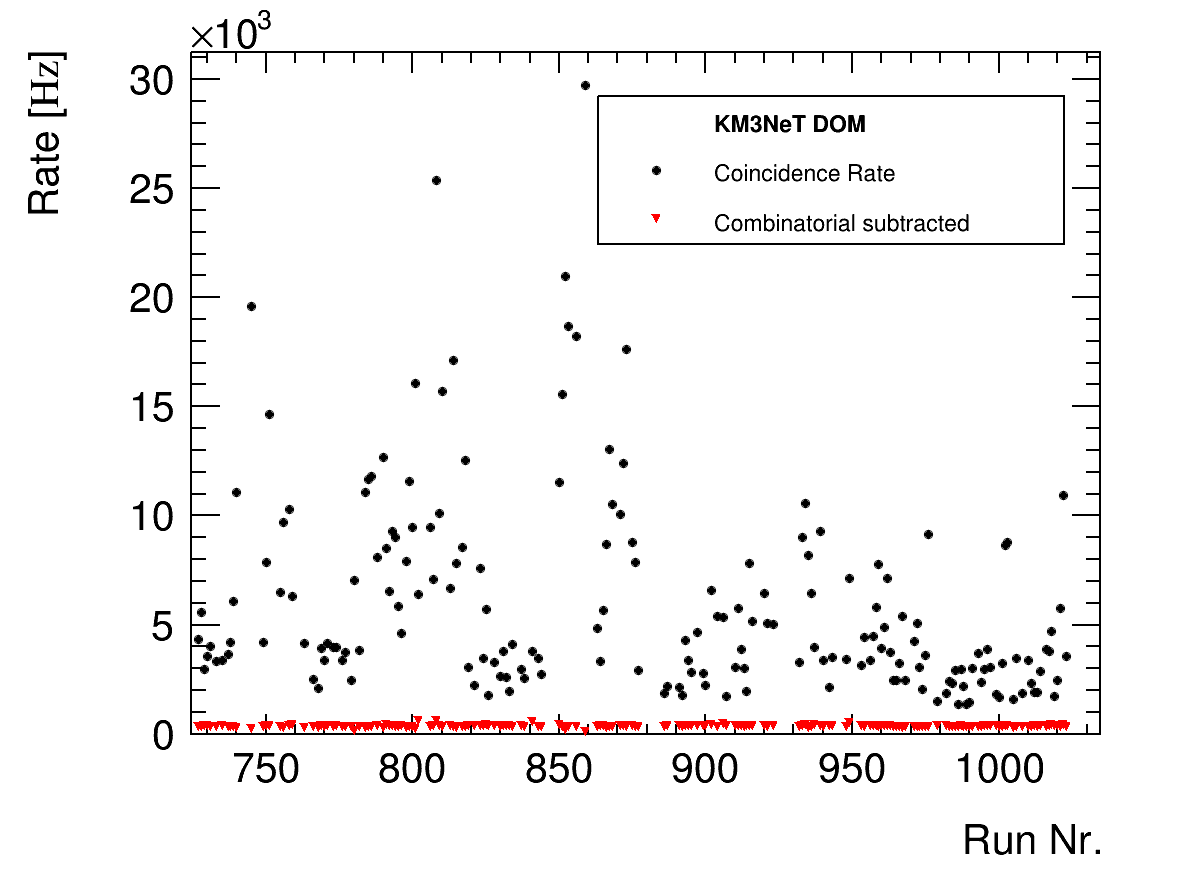}}
\put(30,150){ \includegraphics[width=0.4\textwidth]{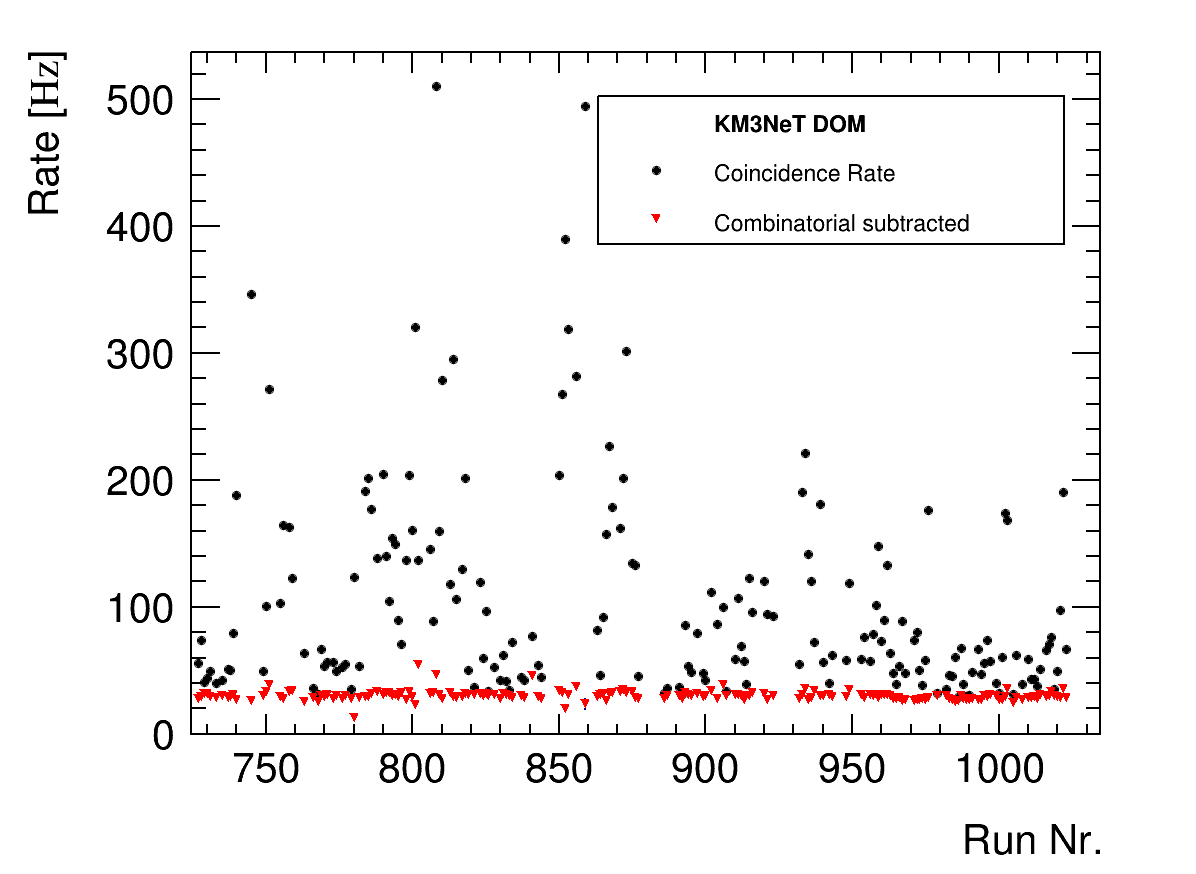}} 
\put(30,0){ \includegraphics[width=0.4\textwidth]{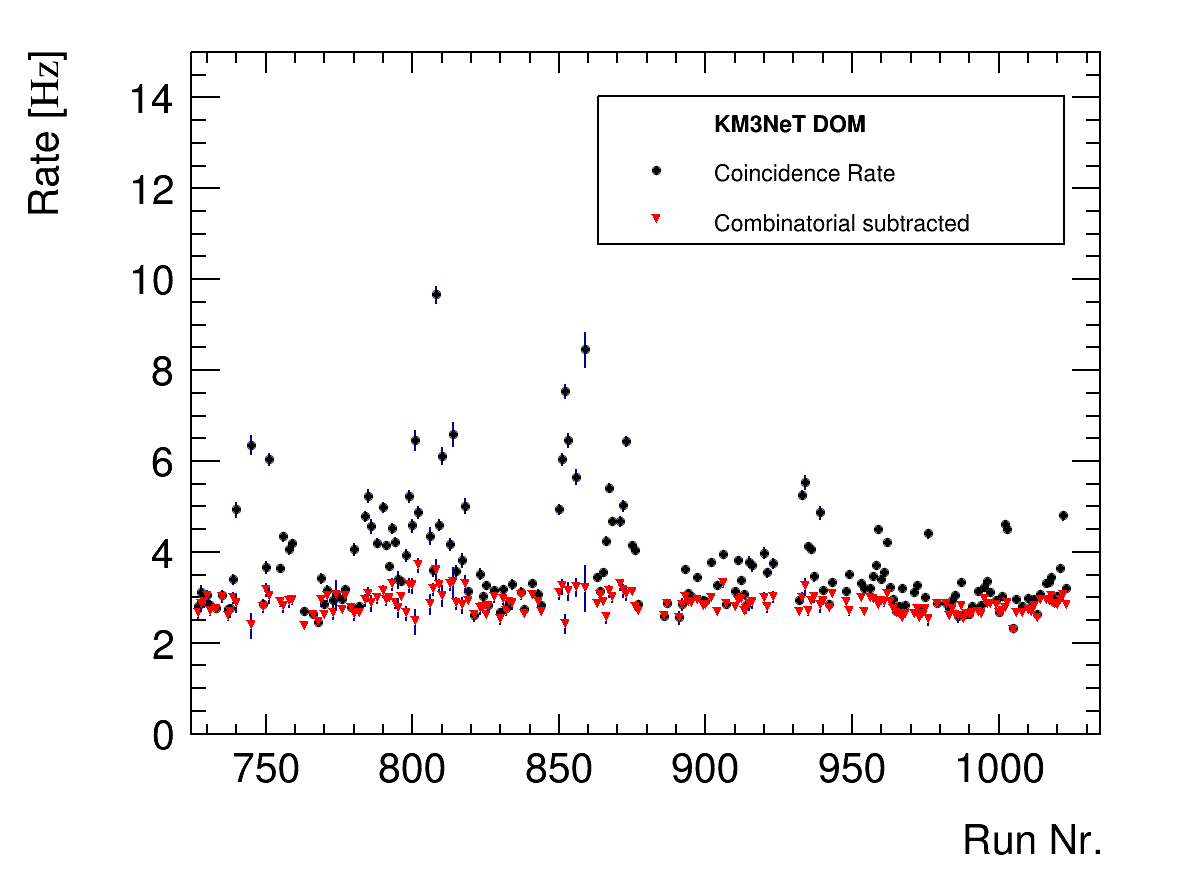}} 
\put(190,550) { \makebox(0,0)[l]{\bf (A)} }
\put(190,395) { \makebox(0,0)[l]{\bf (B)} }
\put(190,245) { \makebox(0,0)[l]{\bf (C)} }
\put(190,95) { \makebox(0,0)[l]{\bf (D)} }
\end{picture}
 \caption{Aggregate rates as a function of run number for (A) singles, (B) twofold coincidence, (C) threefold coincidence and (D) fourfold coincidence rates. The lower red points show the rates with combinatorial background subtracted and gives the true coincidences from $^{40}$K.The runs cover the period from July to December 2013.}
\label{f:runrates}
\end{figure}
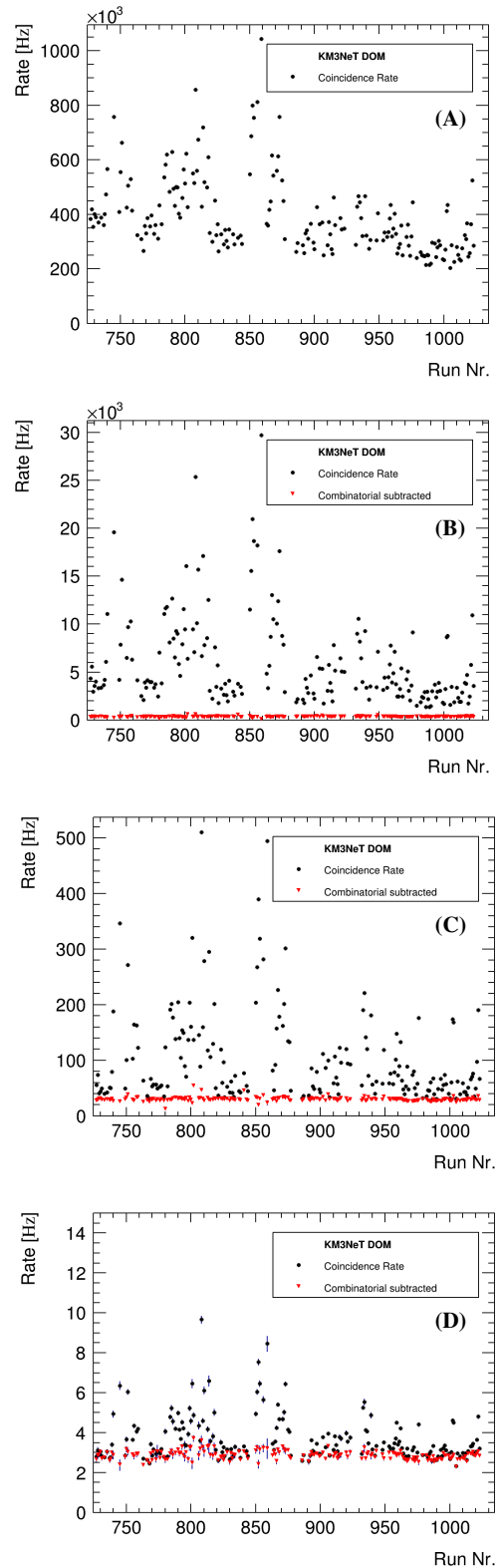
Seawater contains potassium at the level of
416 ppm (at the ANTARES site) of which 0.0118~\% is the radiactive isotope $^{40}$K\ ($\tau_{1/2}$ = 1.28 $\times$ 10$^9$ years). The decay of $^{40}$K yields either $\beta$-electrons or $\gamma$-rays with energies
of approximately \unit{1}{MeV}. Electrons, either directly produced or from Compton scattering, induce Cherenkov radiation while also undergoing multiple Coulomb scattering in the surrounding water. These processes altogether constitute an isotropic source of about 100 detectable photons. These are
the main cause of the baseline singles rate in the PMTs. Assuming a random singles distribution this would yield a baseline twofold coincidence rate of \unit{1200}{Hz} whereas the measured rate is around  \unit{1600}{Hz}. The reason is that the DOM is also capable of detecting multiple photons from a single $^{40}$K decay. Figures~\ref{f:times}A--D show the rates as a function of time difference for increasing angular separation of the PMTs. The clear Gaussian peak centered 
at a time difference of  zero indicates the detection of two photons from the same $^{40}$K decay.
The width of the peak is \unit{6}{ns} (FWHM) corresponding to a single PMT time resolution of $\sigma = \unit{1.8}{ns}$. This signal provides an excellent intra-DOM timing calibration.
The peak becomes less prominent as the angular separation increases 
from 33\degree\ in figure~\ref{f:times}A to 65\degree\ in figure~\ref{f:times}B to 120\degree\  in figure~\ref{f:times}C. When the PMTs are positioned back to back as in figure~\ref{f:times}D the peak has virtually disappeared
leaving only random coincidences. Figure~\ref{f:times2} shows the background subtracted coincidence rate as a function of the angular separation between the PMTs, together with the results of a simulation of the $^{40}$K decay~\cite{b:simuK40}. In general there is good agreement between data and simulation, although the rate at small angular separation is overestimated by the simulation at the 10\% level whereas in the data some random coincidences appear at large separation. \\  
The average total rates observed in the DOM as a function of run number are given in figure~\ref{f:runrates}, without the high rate veto. Each run lasts approximately 10 minutes and the figure spans a period of six months starting in July 2013. The singles rate in figure~\ref{f:runrates}A shows large variations due to the varying bioluminescence component present in the run with an  observed maximum rate of \unit{1.2}{MHz} for the full DOM. As the coincidence level increases, the fluctuations decrease. In general, the  rates are observed to decrease with increasing run number corresponding to a decrease in bioluminescent activity from July through December. Through shifting of each PMT time by 100~ns multiplied by its internal PMT index (1-31), the instantaneous combinatorial background can be measured. In figures~\ref{f:runrates}B--D the lower red curve shows the coincidence rates with this combinatorial background subtracted. These curves thus show the evolution of the genuine $^{40}$K coincidences. As expected, the coincidence rates are completely stable at 340 Hz, 30 Hz and 2.7 Hz for twofold, threefold and fourfold coincidences, respectively. At a coincidence level of four little contribution from bioluminescence is observed.  
\begin{figure}[tp]
\begin{picture}(250,370)
\put(-5,180){ \includegraphics[width=0.5\textwidth]{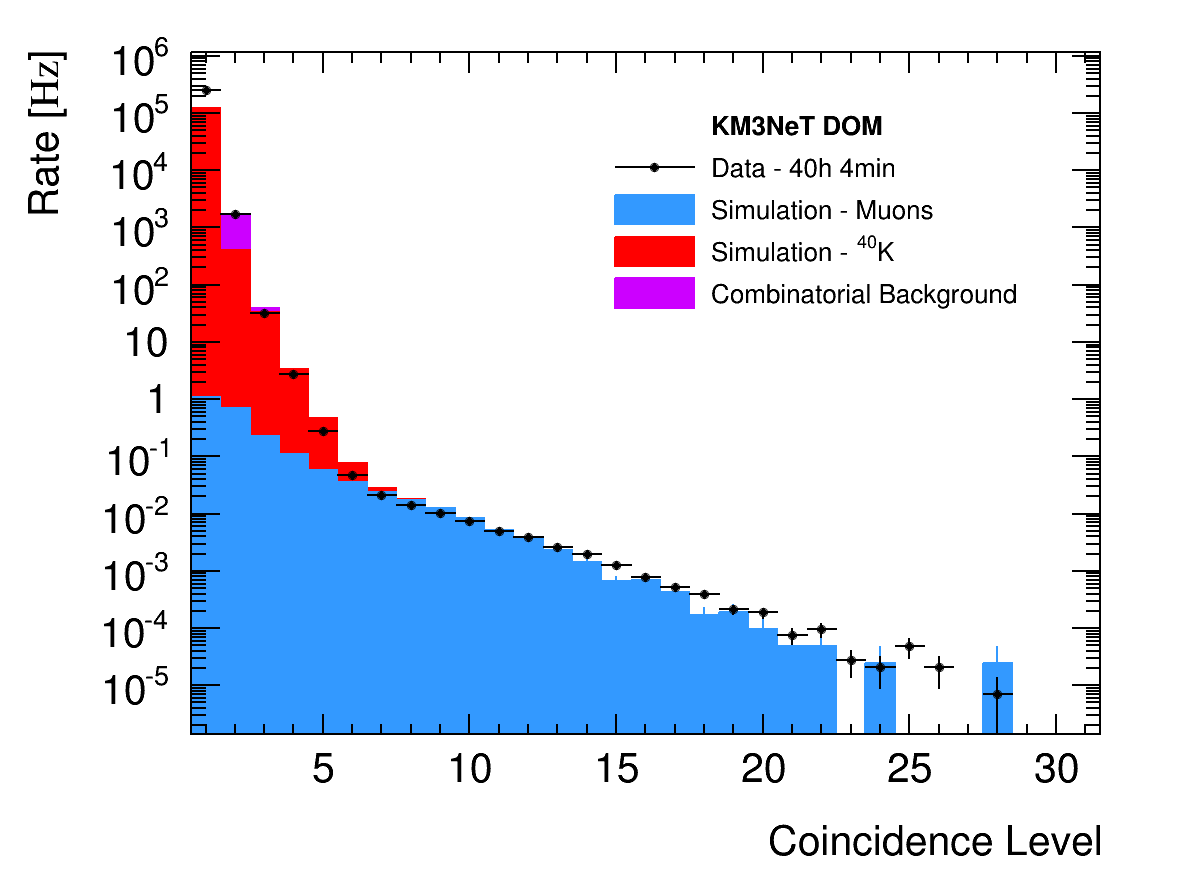}} 
\put(-5,0){  \includegraphics[width=0.5\textwidth]{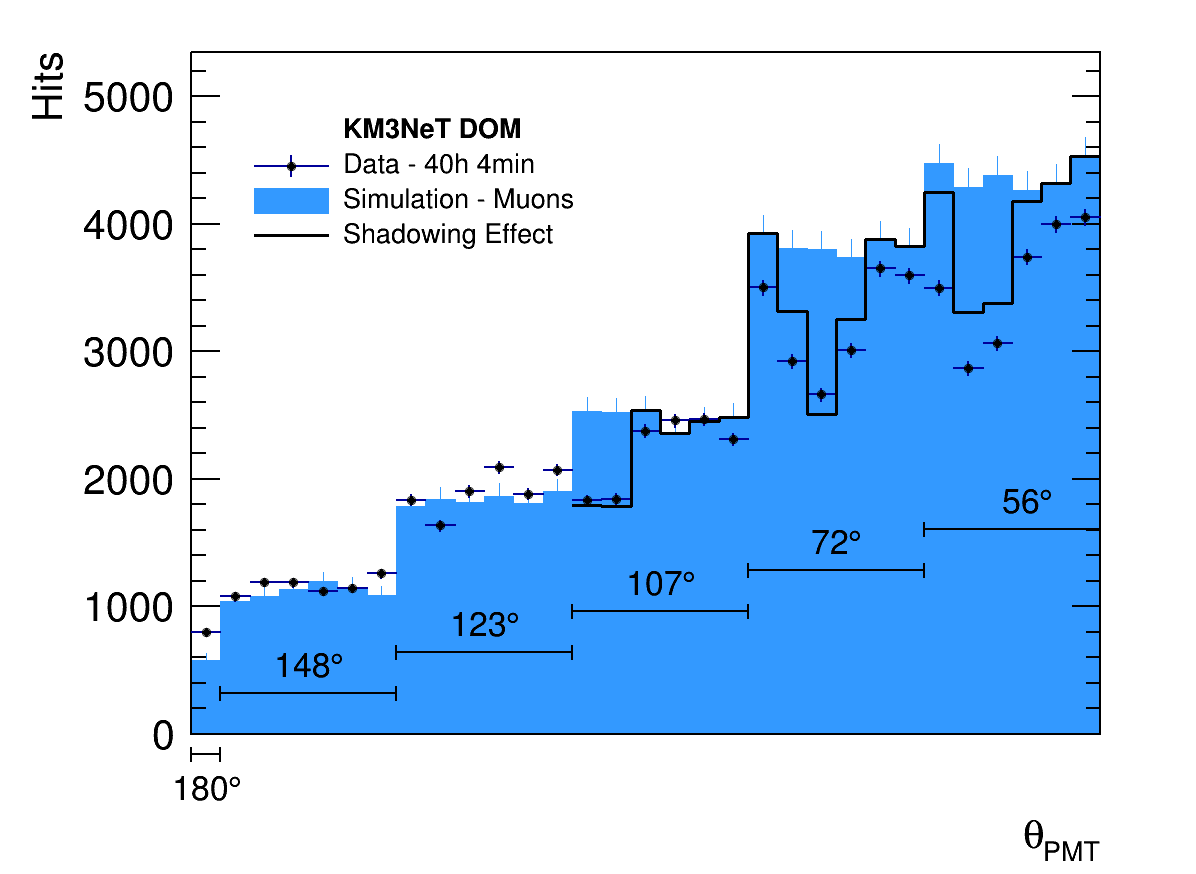} } 
\put(150,290) { \makebox(0,0)[l]{\bf (A)} }
\put(150,150) { \makebox(0,0)[l]{\bf (B)} }
\end{picture}
  \caption{(A) The rate of events as a function of the
    coincidence level (number of PMTs with signal in a $\unit{20}{ns}$
    time window). Black dots correspond to data while coloured
    histograms represent simulations (muons in blue, $^{40}$K in red and
    accidental coincidences in purple). (B) The number of hits
    as a function of the zenith position of the centre of the PMT for coincidence
    levels above seven. One PMT is looking downward ($180^\circ$). The
    others are grouped by six at five different angles. The black dots are data, the blue histogram is   simulation of atmospheric muons and the black histogram show the calculated effect of the shadowing by the ANTARES electronics cylinder. }\label{f:ratesT}
\end{figure}
\section{Analysis}
 The photon counting capability and the directionality
provided by the photocathode segmentation will enable a single DOM to identify  muons and
to be sensitive to their arrival directions, as demonstrated in figure~\ref{f:ratesT}.\\In  figure~\ref{f:ratesT}A, the event rate is shown as a function of the
coincidence level.  The coincidence level is defined as the number of PMTs having  a detected
hit within a $\unit{20}{ns}$ time window.  Below a coincidence level
of six, the measured event rate is in good agreement with the event rate
given by the simulation of the $^{40}$K decays~\cite{b:simuK40}.
 The singles rate (coincidence level 1) is sensitive to the attenuation length in water and also has a contribution from bioluminescence, that is not simulated.
This extra singles rate also gives an additional combinatorial contribution to the twofold coincidences. At the higher coincidence levels the simulation is in good agreement with the data. The rates decrease rapidly as a function 
of the coincidence level as every level increase leads to an extra factor in the acceptance of $\mathrm{A_{PMT}}/4\pi d^2$, where d is the distance form the $^{40}$K decay to the PMT and $\mathrm{A_{PMT}}$ is the sensitive area of the PMT. This leads to a rapidly decreasing volume of water inside which the DOM 
is sensitive to the $^{40}$K decays and therefore a rapid decrease in rate.
 Above the coincidence level of seven, the signals from atmospheric muons  dominate. The simulation of the atmospheric muons was performed using a parameterisation of the measured muon flux and multiplicity for the 2375 m depth of the DOM~\cite{b:simuMuon}.   An excellent agreement is seen between data and simulation
of atmospheric muons. Therefore, with a single DOM muons are unambiguously identified using coincidences of only eight PMTs.\\ 
In figure~\ref{f:ratesT}B, the number of hits detected by each PMT is
shown as a function of their position in terms of zenith angle, corresponding to the rings of PMTs in the DOM. For this figure a cut was applied at a coincidence level larger than seven, which selects a pure
muon sample. With the decrease in zenith angle of the PMT, the rate increases, since atmospheric muons come from above. There is good agreement between data and the atmospheric muon simulation. 
The drop in hit counts for PMTs in the upper three rings is due to a shadowing effect of the
electronics cylinder of the ANTARES line. Since the final KM3NeT suspension of the DOM will not cause such shadowing , incorporating the effect in the Monte Carlo was not considered. Instead the loss in efficiency for the PMTs was calculated assuming all muons arrive exactly vertically and are uniformly distributed horizontally. The black histogram in figure~\ref{f:ratesT}B shows that this first order estimate reproduces the effect quite reasonably.  
\section{Conclusion}\label{s:conc}
The novel digital optical module designed with large photocathode area segmented by the use of 31 small PMTs has been tested in  
deep sea. The DOM was  connected to the instrumentation line of the ANTARES telescope at a depth of around \unit{2375}{m}. Data taking 
was possible continuously from the moment it was connected and is still on-going. The baseline counting rates are stable at around \unit{8}{kHz} per PMT with rates averaged over 10 minutes observed as high as  \unit{1.2}{MHz} for the full DOM. The directional capabilities of the DOM have allowed for the ANTARES support structure to be identified as a cause of excitation of bioluminescent activity. 
The signals from $^{40}$K decay have been used to show the good timing and directional capabilities of the DOM. These signals will provide a straightforward intra-DOM timing calibration in KM3NeT.
With the single DOM it was possible at the level of eightfold coincidences to select a virtually background free sample of atmospheric muons, demonstrating the background suppression 
capabilities of the new design. Finally using the pure muon sample it was possible to demonstrate the sensitivity of the DOM to the arrival direction of the muons.
This design  provides a highly capable optical module for  future neutrino telescopes.
\section*{Acknowledgements}
The authors wish to thank the ANTARES collaboration for agreeing to the installation of the DOM in the detector and for the technical support given.\\
The research leading to these results has received funding from the European Community Sixth Framework Programme under Contract  011937
and the Seventh Framework Programme  under Grant Agreement  212525.


\clearpage

\end{document}